\documentclass[doublespacing]{elsart}

\usepackage{graphicx}
\usepackage{amssymb}

\begin{document}
\begin{frontmatter}

\journal{SCES'2001: }

\title{ESR study of $\bf (Sr,La,Ca)_{14}Cu_{24}O_{41}$}

\author[Koeln]{V.~Kataev\thanksref{permanentaddress}\corauthref{1}}
\author[Koeln]{K.-Y.~Choi\thanksref{newaddress}}
\author[Koeln]{M.~Gr\"uninger}
\author[Koeln,Paris]{U.~Ammerahl}
\author[Koeln]{B.~B\"uchner\thanksref{newaddress}}
\author[Koeln]{A.~Freimuth}
\author[Paris]{A.~Revcolevschi}

\thanks[permanentaddress]{on leave from Kazan Physical Technical Institute, RAS,
Kazan, Russia}
\thanks[newaddress]{currently at: 2. Physikalisches Institut, RWTH Aachen, 52056 Aachen, Germany.}
\corauth[1]{Corresponding author. E-mail: kataev@ph2.uni-koeln.de}

\address[Koeln]{II. Physikalisches Institut, Universit\"{a}t zu K\"{o}ln,
50937 K\"{o}ln, Germany}
\address[Paris]{Laboratoire de Chimie des Solides,
Universit\'e Paris-Sud, 91405 Orsay, France}

\begin{abstract}

We report an electron spin resonance (ESR) study of single crystals of the spin-chain
spin-ladder compound $\rm (Sr,La,Ca)_{14}Cu_{24}O_{41}$. The data suggest that in
intrinsically hole doped Sr$_{14-x}$Ca$_x$Cu$_{24}$O$_{41}$ only a {\em small}
amount of holes is transferred from the chains to the ladders with increasing $x$,
resulting in a crossover from spin dimerized to uniform spin chains. In the
samples of La$_{14-x}$Ca$_x$Cu$_{24}$O$_{41}$ with reduced hole content a very
broad signal is observed in the paramagnetic state, indicative of a
surprisingly strong anisotropy of the nearest neighbor exchange in the chains.

\end{abstract}


\end{frontmatter}

Spin ladders have recently attracted much attention in particular due to
superconductivity with $T_c\!\approx\! 10$ K observed under pressure in $\rm
Sr_{14-x}Ca_{x}Cu_{24}O_{41}$ (SCCO) with $x\geq 11.5$ \cite{Uehara}. SCCO
contains 2-leg $S=1/2$ $\rm Cu_2O_3$ ladders showing a large spin gap $\Delta_{\rm
ladder}\!\approx\! 400$ K \cite{ladderspingap} and $S=1/2$ $\rm CuO_2$ chains,
both running along the $c$-axis. It is 'self-doped' with 6 holes per formula unit.
For $x=0$ almost all holes reside in the chains and show quasi-2D order
\cite{Carter,spindimer}. In this charge ordered state spin dimers with a gap
$\Delta_{\rm dimer}\!\approx\! 130$ K are formed between next-nearest-neighbor Cu
spins via a localized hole \cite{Carter,spindimer}. The conductivity of SCCO
increases with $x$. The prevailing viewpoint is that the chemical pressure due to
Ca doping causes a substantial hole transfer from the chains to the ladders
\cite{Osafune}, i.e., both metallic conductivity and superconductivity are
confined to the ladders. However, recent x-ray absorption data indicate only a
marginal increase of the hole content in the ladders with increasing $x$
\cite{Nuecker}.

We measured electron spin resonance (ESR) of $\rm Cu^{2+}$ ions in single crystals
of $\rm (Sr,\-La,\-Ca)_{14}\-Cu_{24}\-O_{41}$ at 9.47 GHz. A single resonance line
of a Lorentzian shape with an anisotropic $g$-factor is observed. The principal
values of $g$ determined for the $b$- and $c$-axes are $g_{\rm b}=2.29\pm 0.05$
and $g_{\rm c}=2.05\pm 0.03$, respectively. Since $\Delta_{\rm ladder}$ is large,
the ESR signal at $T<300$ K is attributed to the chains. The spin susceptibility
$\chi^{\rm spin}_{ESR}$ derived from the ESR intensity is similar to the static
susceptibility $\chi_{\rm stat}$. In SCCO $\chi^{\rm spin}_{\rm ESR}$ and
$\chi_{\rm stat}$ for small $x$ are well described by the spin-dimer model. A
consistent description of $\chi_{\rm stat}$ in SCCO {\em for all $x$} is obtained
\cite{Udo_sus,PRB} in terms of a {\em smooth crossover} from spin dimers to a
uniform antiferromagnetic (AF) spin chain with increasing Ca content, which
assumes a {\em small} reduction of the hole content in the chains from $\sim\!6$
to $\sim\! 5$. This scenario is justified by the following analysis of the ESR
line width $\Delta H$ (Fig.\ 1). For SCCO with small $x$ (Fig.\ 1, left panel) the
curves above 30 K can be well approximated as $\Delta H(T)= \Delta H_0 + \Delta
H^*(T)$. $\Delta H_0$ is due to $T$-independent spin-spin interactions and
inhomogeneities. $\Delta H^*(T)$ accounts for other, $T$-dependent spin-relaxation
mechanisms.  For $x=0$ it emerges as a strong and almost linear in $T$
contribution to $\Delta H$ only above a certain temperature $T^*\approx 200$ K.\@
$T^*$ decreases rapidly with Ca doping to 170 K for $x=2$ and to $\sim 80$ K for
$x=5$. For still larger $x$ it is not possible to define a $T^*$ below which
$\Delta H$ is constant. We identify $T^*$ as the charge ordering temperature,
consistent with the results of other techniques \cite{Carter,spindimer,Udo_sus}.
Activated hole motion in the chains above $T^*$ breaks spin dimers, causes spin
flips and thus broadens the ESR signal. A rapid reduction of $T^*$ with increasing
$x$ indicates a growing instability of the spin dimer state. Remarkably,  a strong
increase of $\Delta H(T)$ is found for all $x$, suggesting the presence of a
substantial amount of mobile holes in the chains even at large Ca doping. A
consecutive reduction of the total number of holes due to heterovalent
substitution from $n=6$ in SCCO to $n=5$, 4, 3, and 1, as in $\rm Sr_{13}La_1$-,
$\rm La_2Ca_{12}$-, $\rm La_3Ca_{11}$- and $\rm La_5Ca_{9}Cu_{24}O_{41}$,
respectively, results in a much weaker increase of $\Delta H$ with $T$. The slopes
$d(\Delta H)/dT$ are summarized in Fig. 2. According to Ref.\ \cite{Nuecker} all
holes reside in the chains in the compounds with reduced hole doping. Therefore,
the comparison of $d(\Delta H)/dT$ in the left and right part of Fig. 2 gives an
{\em upper border} of approximately {\em one hole} being transferred from the
chains to the ladders in fully hole doped SCCO.

The steep increase of $\Delta H$ at $T<10$ K in SCCO with large $x$ is due to the
development of AF order in the chains. A detailed analysis of the data \cite{PRB}
shows that the increase of the Ca content continuously drives the system towards
an AF instability. This is expected, if spin dimerized chains gradually transform
into uniform chains which eventually order at low $T$ due to weak interchain
couplings.

With increasing $x$ the $T$-independent part of the line width $\Delta H_0$ in
SCCO increases by a factor of 10. In the samples with reduced hole doping it grows
up to $\sim\! 1.5$ kOe in nearly undoped $\rm La_5Ca_{9}Cu_{24}O_{41}$. A careful
analysis of the spectra suggests that this effect is not related to structural or
magnetic inhomogeneities \cite{PRL}. At $T\! <\! 50$ K the resonance line
experiences additional broadening when long-range AF order at $T_N\! \approx\! 10$
K with an extreme magnetic anisotropy is approached \cite{AmmerahlIsing}. The
width of the ESR signal in the {\em paramagnetic} regime above $\sim\! 50$ K,
where static short-range correlations vanish, is therefore determined mainly by
the anisotropy of the spin-spin interactions, which in concentrated paramagnets is
the major broadening mechanism. The dominant isotropic exchange in the chains with
a small hole content is ferromagnetic and occurs between {\em nearest} neighbors,
$H_{\rm iso}\! =\! J_{\rm iso}\sum {\bf S_i}{\bf S_{i+1}}$, with $J_{\rm iso}\!
\approx\! -20$ K \cite{Carter}. The leading anisotropy is of a symmetric type,
$H_{\rm aniso}\! =\! \sum {\bf S_i}A_{\rm i,i+1}{\bf S_{i+1}}$. A conventional
estimate of $A_{\rm i,i+1}$ with $\Delta H\!\sim\! 1.5$ kOe yields $A_{\rm
i,i+1}\! \sim\! 2$ K \cite{PRL}, which is as large as 10\% of $J_{\rm iso}$. Such
a strong anisotropy is surprising for copper oxides, which are considered to be
the best experimental realizations of an isotropic Heisenberg magnet. It can be
explained by the specific geometry of two symmetrical 90$^\circ$ Cu--O--Cu bonds,
connecting nearest neighbor Cu sites. In this geometry the influence of the
spin-orbit coupling on the superexchange is found to be considerably enhanced
\cite{theory}. The ESR data indeed show that low-dimensional cuprates with certain
bonding geometries may deviate significantly from the isotropic Heisenberg model.

\begin{figure}
     \centering
     \includegraphics[width=75mm]{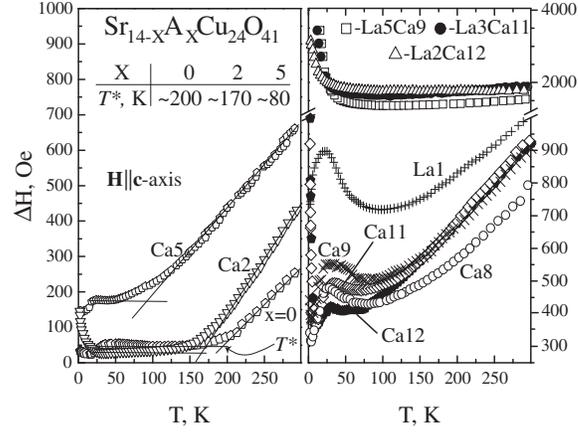}
     \caption{$T$-dependence of the ESR line width $\Delta H$ of $\rm (Sr,La,Ca)_{14}Cu_{24}O_{41}$.
     Solid lines in the left
     panel denote the constant and
     linear contributions to $\Delta H(T)$, as explained in the text. $T^*$ indicates the charge ordering temperature. }
\end{figure}

\begin{figure}
     \centering
     \includegraphics[width=75mm]{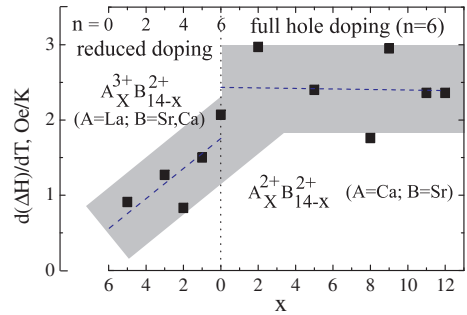}
     \caption{High temperature slope of the $\Delta H(T)$ dependence
     of the samples with full ($n=6$) and reduced ($n<6$) hole doping.}
\end{figure}

\end{document}